# Origin of nonlinear photocurrents in chiral multifold semimetal CoSi unveiled by terahertz emission spectroscopy

Yao-Jui Chan,[1] Syed Mohammed Faizanuddin,[1,2,3] Raju Kalaivanan,[1] Sankar Raman,[1] Hsin Lin,[1] Uddipta Kar,[1] Akhilesh Kr. Singh,[1] Wei-Li Lee,[1] Ranganayakulu K. Vankayala,[1] Min-Nan Ou,[1] and Yu-Chieh Wen[1,*]

[1]*Institute of Physics, Academia Sinica, Taipei 11529, Taiwan*

[2]*Taiwan International Graduate Program, Academia Sinica, Taipei 11529, Taiwan*

[3]*Department of Engineering and System Science, National Tsing Hua University, Hsinchu 300044, Taiwan*

*Correspondence to: ycwen@phys.sinica.edu.tw (Y.C.W.)

**ABSTRACT**

Spectroscopic identification of distinct nonlinear photocurrents unveils quantum geometric properties of electron wavefunctions and the momentum-space topological structures. This is especially interesting, but still puzzling, for chiral topological semimetals with possibilities of hosting giant quantized circular photogalvanic effect. Here we report a comprehensive terahertz (THz) emission spectroscopic analysis of nonlinear photoconductivity of chiral multifold CoSi at 0.26 ~ 1 eV. We find a large linear shift conductivity (17 μA/V$^2$), and confirm a giant injection conductivity (167 μA/V$^2$) as a consequence of strongly interfered non-quantized contributions from the vicinity of multifold nodes with opposite chiralities. The bulk injection current excited by the pump field with a complex wavevector is shown to carry both longitudinal and transverse components. Symmetry analyses further unveil weak nonlocal photon drag effect in addition to the photogalvanic effect. This work not only highlights chiral transition metal monosilicides for mid-infrared photovoltaic applications via various nonlinear optical channels, but also consolidates the THz spectroscopy for quantitative photovoltaic research.

# I. INTRODUCTION

Bulk photovoltaic effect being a second-order nonlinear optical effect rectifies light into a dc photocurrent in homogeneous non-centrosymmetric materials [1-4]. Its local response, known as the photogalvanic effect (PGE), originates from the inversion-asymmetric transition of electron position or velocity during the optical excitation. The resulting photocurrents are called shift current and injection current, respectively [4,5]. Besides the PGE, the nonlocal photon drag effect (PDE) can also yield photocurrents involved with momentum transfer from photons to electrons [6,7]. These processes are dictated by symmetry and quantum geometric properties of the electron Bloch wavefunction [4,7-12], and can thus be largely enhanced near the topological nodes in semimetals through, for example, the Berry curvature dipole and shift current dipole [7,10]. The effects offer promising opportunities to develop sensitive and broadband infrared/THz photodetectors at room temperature [3,9,13]. Here the dark current noise is suppressed since the mechanism does not need an external or internal bias, in contrast to the conventional diode- or junction-based photodetection scheme.

Weyl semimetals possessing no inversion and mirror symmetries is especially interesting. de Juan *et. al.* proposed that the circular photogalvanic effect (CPGE) therein is large and effectively quantized in terms of the chiral charge of Weyl points without material-dependent parameter [8]. Experimental search for this phenomenon were soon conducted by Rees *et. al.* [14] and Ni *et. al.* [15,16] using chiral multifold (Rh,Co)Si system through THz emission spectroscopy (TES), but the results were inconclusive in three aspects. First, the nonlinear photoconductivity, $\overleftrightarrow{\sigma}^{(2)}$, measured by the two groups (for RhSi) differed largely by a factor of ~20 [14,15]. The discrepancy may arise from varying crystal qualities, while different approaches in processing the TES data might appear as another concern. Secondly, the photon drag effect was simply ignored in the analysis. Thirdly, and most importantly, the plateau-like spectral response of the quantized CPGE was not consistently identified [14,15]. Whether the undoped crystals host the quantization remains a hot debate [11,12]. We aim to examine these issues in this paper.

Spectroscopic identification and quantification of nonlinear photocurrents are key ingredients in exploring the topology-derived PGE/PDE and prerequisite to development of advanced photovoltaics. The TES offers a contact-free means for probing ultrafast photocurrents by detecting the accompanying THz radiation. It allows for assessing the intrinsic material properties, in contrast to conventional dc current measurements that would suffer errors from the contact resistance and thermal current and non-local diffusion of photoexcited carriers to the contacts [13,16]. Nevertheless, applications of TES to quantitative measurements of $\overleftrightarrow{\sigma}^{(2)}$ remain challenging and were not fully implemented [14-18].

Here we report an improved THz emission spectroscopic analysis for quantitative characterization of the distinct photocurrents in chiral semimetal CoSi. Results indicate a large linear shift conductivity, and reveal giant longitudinal injection currents appearing as a consequence of strongly interfered non-quantized contributions from the vicinity of multifold nodes with opposite chiralities. Bulk transverse injection currents and relatively weak photon drag effect are also identified. This work not only highlights the B20-type chiral cubic crystals for potential photovoltaic applications but also consolidates TES for topological nonlinear optics research.

**II. EXPERIMET**

The sample under study is a (110) CoSi single crystal prepared by the floating zone growth technique. See Supplementary Materials (SM) section I for details of its growth condition and supports of X-ray diffractions and rotational-anisotropy optical sum-frequency generation for its good crystallinity. Our TES experiment was based on a 1-kHz, 50-fs Ti:sapphire laser equipped with an optical parametric amplifier (OPA), followed by difference frequency generation (DFG). Optical pumps with a bandwidth of ~30 meV came from signal, idler, or their DFG from the OPA, enabling a frequency scanning between 0.26 ~ 1.03 eV. They illuminated the sample with an incident angle $\theta_i = 45°$ and the fluence of ~1 mJ/cm². As sketched in Fig. 1(a), the crystal was set in the (*x,y,z*) lab frame with $[110] \parallel \hat{z}$ and $[001] \parallel \hat{y}$ (unless specified below), with $\hat{z}$ along the surface normal and $\hat{x}$-$\hat{z}$ being the scattering plane. The induced THz radiation was detected by the electro-optic sampling scheme with a 2 mm-thick (110) ZnTe crystal [16,17]. All measurements were conducted in dried air at room temperature with controlled (selected) optical (THz) polarization. For analyses of the deduced spectra, we must first describe the model we adopted for the THz emission from metals.

## III. THEORY

We formulate the THz radiation from nonlinear photocurrents in *opaque* media. This improves the conventional theory of TES by appropriately treating the effects of complex wavevectors on Fresnel transmission and phase matching [14-18]. Besides quantitative accuracy, we shall demonstrate its capability of predicting the bulk transverse CPGE current, a feature omitted previously, in the B20-type chiral cubic crystal [16,19].

We consider an optical pulse with its field $\vec{E}^m = \mathcal{E}^m \hat{e}^m$ at central frequency $\omega$ incident from air onto a solid [Fig. 1(a)]. Here $\mathcal{E}$ is the field amplitude, $\hat{e}$ is the polarization unit vector, and the superscript used here and below denotes the medium ($m = I, II$ for air and solid, respectively). The light field generates an ultrafast current $\vec{J}^{(2)}$ at the frequency $\omega_{THz}$ via an intra-pulse DFG process, as described by

$$\vec{J}^{(2)}(\omega_{THz}) = \overleftrightarrow{\sigma}^{(2)}:\vec{E}^{II}(\omega+\omega_{THz})\vec{E}^{II^*}(\omega) \cong \overleftrightarrow{\sigma}^{(2)}:\vec{E}^{II}(\omega)\vec{E}^{II^*}(\omega). \qquad (1)$$

With $d\vec{J}^{(2)}/dt$ being a source term in Maxwell's wave equation, one can follow Ref. [20,21] to derive the radiated THz field in air, $\vec{E}^I(\omega_{THz})$, by integrating $\vec{J}^{(2)}$ over the depth in the solid with the phase mismatch $\Delta k_z^{II}$ and the Fresnel transmission coefficients of the interface $\overleftrightarrow{L}$ taken into account. Amplitude of the radiation is expressed by

$$\mathcal{E}_{\zeta\xi}^I(\omega_{THz}) = f\left\{[\overleftrightarrow{L}(\omega_{THz}) \cdot \hat{e}_\xi^I(\omega_{THz})] \cdot \frac{\overleftrightarrow{\sigma}^{(2)}}{\Delta k_z^{II}} : [\overleftrightarrow{L}(\omega) \cdot \hat{e}_\zeta^I(\omega)][\overleftrightarrow{L}(\omega) \cdot \hat{e}_\zeta^I(\omega)]^*\right\} |\mathcal{E}^I(\omega)|^2, \qquad (2)$$

with $f = -(2c\sqrt{2\epsilon_0 c}\cos\theta_e)^{-1}$, $\epsilon_0$ the vacuum dielectric constant, $c$ the vacuum light speed, and $\theta_e (= \theta_i)$ the THz emission angle. $\zeta$ and $\xi$ denote the polarization states of the optical and THz fields, respectively.

We remark on the boundary condition in calculating $\Delta k_z^{II}$ and $\overleftrightarrow{L}$. As sketched in Fig. 1(b), with a real-valued wavevector $\vec{k}^I$ of the incident wave, the transmitted wave in the absorbing solid must have a complex wavevector $\vec{k}^{II} = \vec{\alpha}^{II} + i\vec{\beta}^{II} = k_0(n_\alpha \hat{\alpha}^{II} + in_\beta \hat{\beta}^{II})$ in order to preserve continuity of the tangential electric fields in the two media at the interface [22,23]. Here $k_0 \equiv 2\pi/\lambda$ with $\lambda$ the vacuum wavelength. The unit vectors $\hat{\alpha}^{II}$ and $\hat{\beta}^{II}$ depict the normals of the planes with constant phase and constant amplitude, respectively. They do not coincide in lossy media in general due to complex dielectric constant $\tilde{\epsilon}^{II}$. $n_\alpha$ and $n_\beta$ can be interpreted as *apparent* refractive indices, which relate to $\tilde{\epsilon}^{II}$ through $\vec{k}^{II} \cdot \vec{k}^{II} = k_0^2 \tilde{\epsilon}^{II}$ [22,23]. Together with the continuity boundary condition, one can solve $\vec{k}^{II}$ from a given $\vec{k}^{I}$ for optical or THz field, and then use $\vec{k}^{II}$ and $\vec{k}^{I}$ to calculate the Fresnel transmission coefficients $\overleftrightarrow{L}$ and the complex phase mismatch $\Delta k_z^{II} = k_z^{II}(\omega + \omega_{THz}) - k_z^{II}(\omega)^* - k_z^{II}(\omega_{THz})$ with $k_z \equiv \hat{z} \cdot \vec{k}$. (See SM section II for the solutions of $\vec{k}^{II}$ and the expression of $\overleftrightarrow{L}$.) For CoSi, the results so obtained with its previously measured $\tilde{\epsilon}^{II}$ [16] are displayed in Fig. S3 in SM.

Regarding the nonlinear photocurrent, it can be expressed as a Taylor expansion of Eq. (1) with respect to the real part of $\vec{k}^{II}$:

$$J_l^{(2)} = \sigma_{lmn}^{(2)}(\omega, \vec{\alpha}): E_m E_n^* = \mu_{lmn}^{(2)}(\omega): E_m E_n^* + v_{lmnq}^{(2)}(\omega): E_m E_n^* \alpha_q + \mathcal{O}(\alpha^2) + \cdots. \quad (3)$$

where (*l*, *m*, *n*, *q*) are referred to the Cartesian components, and repeated indices are implicitly summed over. We omit the superscript *II* for clarity. $\overleftrightarrow{\mu}^{(2)}$ and $\overleftrightarrow{v}^{(2)}$ that characterize the local and the first-order non-local responses of $\overleftrightarrow{\sigma}^{(2)}$ are the PGE and PDE coefficients, respectively [2,7]. By approximating $\vec{J}^{(2)}$ as a dc current, one can further split $\vec{J}^{(2)}$ into two parts with respect to the photon angular momentum: $\vec{J}^{(2)} = \vec{J}^L + \vec{J}^C$ with $J_l^L \equiv \text{Re}\sigma_{lmn}^{(2)} \cdot (E_m E_n^* + E_n E_m^*)/2$ and $J_l^C \equiv i\text{Im}\sigma_{lmn}^{(2)} \cdot (E_m E_n^* - E_n E_m^*)/2$. $\vec{J}^L$ ($\vec{J}^C$) contributed from $\text{Re}\overleftrightarrow{\sigma}^{(2)}$ ($\text{Im}\overleftrightarrow{\sigma}^{(2)}$) is referred to as linear (circular) PGE/PDE current since it is proportional to the Stokes parameter $S_2$ ($S_3$) that is sensitive to linearly (circularly) polarized light [2]. For CoSi of space group 198, there are five independent non-zero $v_{lmnq}^{(2)}$ elements [24] and one for the PGE: $\mu_{lmn}^{(2)} = \rho|\epsilon_{lmn}| + i\eta\epsilon_{lmn}$ with $\epsilon_{lmn}$ the Levi-Civita symbol and *l*, *m*, *n*, *q* $\in$ (*a*,*b*,*c*) in the crystal coordinates. The linear (L-) PGE conductivity $\rho$ and CPGE conductivity $\eta$ characterize the linear shift and circular injection currents in non-magnetic CoSi, respectively [4]. With above, the constituted $\overleftrightarrow{\sigma}^{(2)}$ in the crystal coordinates is expressed as

$$\overleftrightarrow{\sigma}^{(2)} = \begin{pmatrix} \begin{pmatrix} v^{(2)}_{aaaa}\alpha_a \\ v^{(2)}_{bbcc}\alpha_b \\ v^{(2)}_{ccbb}\alpha_c \end{pmatrix} & \begin{pmatrix} v^{(2)*}_{bbcc}\alpha_b \\ v^{(2)}_{bccb}\alpha_a \\ \rho + i\eta \end{pmatrix} & \begin{pmatrix} v^{(2)*}_{ccbb}\alpha_c \\ \rho - i\eta \\ v^{(2)}_{cbbc}\alpha_a \end{pmatrix} \\ \begin{pmatrix} v^{(2)}_{cbbc}\alpha_b \\ v^{(2)*}_{ccbb}\alpha_a \\ \rho - i\eta \end{pmatrix} & \begin{pmatrix} v^{(2)}_{ccbb}\alpha_a \\ v^{(2)}_{aaaa}\alpha_b \\ v^{(2)}_{bbcc}\alpha_c \end{pmatrix} & \begin{pmatrix} \rho + i\eta \\ v^{(2)*}_{bbcc}\alpha_c \\ v^{(2)}_{bccb}\alpha_b \end{pmatrix} \\ \begin{pmatrix} v^{(2)}_{bccb}\alpha_c \\ \rho + i\eta \\ v^{(2)*}_{bbcc}\alpha_a \end{pmatrix} & \begin{pmatrix} \rho - i\eta \\ v^{(2)}_{cbbc}\alpha_c \\ v^{(2)*}_{ccbb}\alpha_b \end{pmatrix} & \begin{pmatrix} v^{(2)}_{bbcc}\alpha_a \\ v^{(2)}_{ccbb}\alpha_b \\ v^{(2)}_{aaaa}\alpha_c \end{pmatrix} \end{pmatrix}. \quad (4)$$

Here the element $\sigma^{(2)}_{lmn}$ is the *n*-th element of the column vector in the *l*-th row and *m*-th column of the matrix. [Also see Eq. (S8) in SM for the lab-coordinate representation.] Besides PGE and PDE, the transient photocurrent can arise from space charge separation during diffusion of the photocarriers along the surface normal. It is known as the photo-Dember effect (PDmE) for *p*-polarized THz generation [25]. We shall examine these three mechanisms below.

**IV. Result and Discussion**

Our investigation starts by inspecting polarization-dependent THz emissions from CoSi at $\hbar\omega = 0.61$ eV. As shown in Fig. 2, the THz radiation varies in strength prominently with respect to the polarization combination. With linearly-polarized optical pumps, the THz field is comparable to our noise floor for *SS* polarization and increases in strength from *SS*, *PP*, to *PS*. (The optical and THz polarizations are labelled in order as *S*, *P*, *R*, or *L* for *s*-, *p*-, right-hand circular, or left-hand circular polarization, respectively.) One can readily compare this observation with the $\overleftrightarrow{\sigma}^{(2)}$ tensor [Eq. (4) or Eq. (S8) in SM] for an intuition about the different mechanisms. Particularly, with $[110] \parallel \hat{z}$, $[001] \parallel \hat{y}$, and $\alpha_c = 0$, the PGE is active only for *PS*, whereas the PDE, similar to the PDmE, contributes to *PP*. The exclusive role of the PGE to the strongest *PS* signal suggests a significant PGE contribution to the photocurrent, in comparison with the PDE- and/or PDmE-induced moderate *PP* THz field. All the three mechanisms are forbidden for *SS* polarization, in line with its imperceptible THz emission in Fig. 2. As using circularly polarized pumps, we find the (*p*-polarized) THz emission to display manifest helicity dependency. This feature was reported previously with a CPGE-derived explanation [14-16], while unknown impacts from the PDE need to be clarified.

For more quantitative insights, we examine the THz emission with the optical pump polarization controlled by a rotating quarter waveplate (QWP). In this case, amplitude of the THz field can be expressed from Eq. (2) and (S8) with $\hat{e}^I(\omega)$ set by the QWP with its fast axis at angle $\psi$ away from the scattering plane, as given by

$$\mathcal{E}^I(\omega_{THz}) \propto A_1 \cdot \cos 4\psi + A_2 \cdot \sin 4\psi + A_3 \cdot \sin 2\psi + A_4, \tag{5}$$

where factors $A_i$ ($i = 1 \sim 4$) are functions of $\overleftrightarrow{L}$ and the PGE/PDE coefficients with their expressions given by Eq. (S10) and (S11) in SM. This derivation turns out a symmetry-dictated selection rule: For *p*-polarized THz emission, $A_2$ and $A_3$ ($A_1$ and $A_4$) originate from the PGE (PDE and PDmE). For *s*-polarized THz emission, however, one has the PGE-induced $A_4$ ($=3A_1$) and the PDE-induced $A_2$ and $A_3$, but the PDmE is forbidden. Experimentally, we measure the THz emission from CoSi as a function of $\psi$ and then fit the deduced $\mathcal{E}^I(\omega_{THz})$ with variables $A_i$ via Eq. (5), such that the contributions of the distinct mechanisms can be decomposed. As shown in Fig. 3, the measured *p*-polarized THz emission exhibits clear $\psi$ dependency with a predominant PGE contribution, i.e., $\mathcal{E}^I(\omega_{THz}) \approx A_2 \sin 4\psi + A_3 \sin 2\psi$. Consistently, we find the *s*-polarized THz emission to be governed by the PGE with $\mathcal{E}^I(\omega_{THz}) \approx A_1 \cos 4\psi + A_4$, whereas a weak PDE response can be identified via the $A_2 \sin 4\psi$ term. Note that we deduce $A_4/A_1 \approx 2.4$, in good agreement with the theoretical prediction (3).

To further confirm the minor role of the PDE, we examine the THz emission with respect to azimuthal rotation of the crystal, with a focus here on *SS* and *PS* polarization for excluding the PDmE. If the PGE overwhelms the PDE, one expects the THz field amplitudes for the two polarization combinations to follow the symmetry of $\overleftrightarrow{\mu}^{(2)}$ and a unique set of $(\rho, \eta)$. Indeed, the measured *SS*– and *PS*–polarized THz emissions appear to be highly anisotropic (Fig. 4) and can be depicted simultaneously well using a single set of fitting variables $(\rho, \eta)$ with the known $\overleftrightarrow{L}(\omega)$ through the PGE-derived expressions [from Eq. (2) and (S8) with $\overleftrightarrow{\sigma}^{(2)} = \overleftrightarrow{\mu}^{(2)}$]:

$$\mathcal{E}^I_{SS}(\omega_{THz}) = B\rho \cdot 6(\cos\phi - \cos^3\phi)|L_{yy}(\omega)|^2, \tag{6a}$$

and

$$\mathcal{E}^I_{PS}(\omega_{THz}) = B\{\rho \cdot [(3\cos^3\phi - 2\cos\phi)|L_{xx}(\omega)|^2 - \cos\phi\,|L_{zz}(\omega)|^2] + \eta \cdot 2\mathrm{Im}[L_{zz}(\omega)L^*_{xx}(\omega)]\}, \tag{6b}$$

where $B$ is a trivial common constant, and $\phi$ is the angle of the crystal axis $[001]$ from $\hat{y}$.

A new insight from Eq. (6b) is regarding an isotropic ($\phi$-independent) CPGE in addition to an anisotropic LPGE. The former cannot be ignored in our data fitting in order to depict the observed offset in Fig. 4(c) and one-fold asymmetric lobes in Fig. 4(d). This *PS*-polarized feature therefore demonstrates a transverse circular injection current excited by a linearly polarized input

– a feature absent in earlier symmetry arguments [16,19]. The reason appears to be simple, as a linear optics problem. For space group 198, one derives $\vec{J}^C = i\eta\left(\vec{E}^{II} \times \vec{E}^{II*}\right)$ and expects a purely longitudinal injection current when approximating $(\vec{E} \times \vec{E}^*) \parallel \vec{k}$ [16,19] It is, however, not the general case for opaque solids. For instance, an incident *p*-polarized optical field will experience unequal phase changes in its $x$ and $z$ components through $\overleftrightarrow{L}$ at the interface due to complex $k_z^{II}$ [Eq. (S4) in SM]. The unbalanced phase between $\hat{x} \cdot \vec{E}^{II}$ and $\hat{z} \cdot \vec{E}^{II}$ leads to $\hat{y} \cdot \left(\vec{E}^{II} \times \vec{E}^{II*}\right) \neq 0$, i.e., finite light helicity perpendicular to $\vec{k}^{II}$, and hence excites a transverse $\vec{J}^C$, as described by Eq. (6b) and demonstrated experimentally. Note that the conventional TES theory fails to capture this phenomenon because $\arg(L_{xx}) \neq \arg(L_{zz})$ was not realized [14-18].

With origins of the photocurrents clarified, we can now revisit the CPGE conductivity spectrum of CoSi for better quantitative insights. To fully exclude the PDE and PDmE in the analysis, we measure the THz radiations with *PS*, *LP*, and *RP* polarization combinations at $\phi = 0°$. In this case, both PDE and PDmE are forbidden by symmetry for *PS* polarization; while they act on $\mathcal{E}_{LS}^I(\omega_{THz})$ and $\mathcal{E}_{RS}^I(\omega_{THz})$, their effects can be eliminated through signal differentiation. A simple derivation from Eq. (2) and (S8) proves

$$\mathcal{E}_{PS}^I(\omega_{THz}) = (D_1 \cdot \rho + D_2 \cdot \eta)|\mathcal{E}^I(\omega)|^2, \qquad (7a)$$

$$\Delta\mathcal{E}^I(\omega_{THz}) \equiv (\mathcal{E}_{LP}^I - \mathcal{E}_{RP}^I)/2 = (D_3 \cdot \rho + D_4 \cdot \eta)|\mathcal{E}^I(\omega)|^2. \qquad (7b)$$

We have lumped $\overleftrightarrow{L}$ and $\Delta k_z^{II}$ into constants $D_i$ ($i = 1\sim4$). [See Eq. (S12) in SM for their expressions.] To quantify the photoconductivity, we follow Ref. [17] to normalize the THz emission spectra of CoSi against a reference (110) ZnTe emitter crystal that yields a THz radiation $\mathcal{E}^{I,ZT}(\omega_{THz}) = \chi^{ZT} D_0 |\mathcal{E}^I(\omega)|^2$ with $D_0(\overleftrightarrow{L}, \Delta k_z^{II})$ expressed in Eq. (S12) in SM. With the nonresonant second-order optical susceptibility $\chi^{ZT} = 107$ pm/V for ZnTe [26] and $D_i$ ($i = 0 \sim 4$) calculated from the known $\tilde{\epsilon}^{II}$ [16,27,28], one can then determine $\eta$ from the measured THz spectra for a given $\hbar\omega$ via $|\eta| = \left|\frac{\chi^{ZT} D_0}{D_1 D_4 - D_3 D_2}\right| \cdot \left|D_1 \frac{\Delta\mathcal{E}^I}{\mathcal{E}^{I,ZT}} - D_3 \frac{\mathcal{E}_{PS}^I}{\mathcal{E}^{I,ZT}}\right|$ [29]. Figure 5 shows the injection conductivity spectrum so obtained with $\hbar\omega = 0.26 \sim 1.03$ eV. It displays a spectral hump at ~0.38 eV with strength up to 167 μA/V². This observation is similar to that reported earlier but somewhat different in its strength and (narrower) lineshape [16].

In explaining the spectrum, we follow Ni *et. al.* [16] to elucidate the effects of the multifold fermions through the density functional theory (DFT)-derived band structure. (See Ref. [16] for the calculation or the inset of Fig. 5 for a sketch.) CoSi features two multifold crossings of the band structure at the Γ and R points with topological charges of ±4 [16,30]. Without spin-orbit coupling (SOC), they take the form of a degenerate spin-1 fermion at Γ near the Fermi level $E_F$ and a spin-degenerate double Weyl crossing at R with energy of -185 meV below $E_F$ [16]. (The moderate SOC splitting in CoSi, e.g., ~20 meV at the Γ point node, is neglected in our discussion.) These multifold fermions contribute to the CPGE in a competitive manor. Without an intentional doping, the transitions near the R point are Pauli blocked for $\omega$ below ~0.25 eV but active for higher $\omega$. Optical transition across the Γ-point node near $E_F$ is also affected by Pauli blocking, especially for these involved with the flat band 2. (We label the three crossing bands at Γ from bottom to top as band 1 to 3.) Particularly, the transition from band 1 to 2 (from band 2 to 3) form an enclosed manifold in the momentum space at $\omega \approx 0.02 \sim 0.1$ eV ($\omega > 0.45$ eV). The complete optical transitions in these spectral windows are expected to induce prominent CPGE contributions from the Γ node, whereas a reduced strength due to incomplete transitions occurs in the frequency gap between. Therefore, the $\eta(\omega)$ spectrum resulting from interference of the two oppositely charged nodes is likely to feature a bipolar lineshape, with dominant Γ−node (R−node) contributions below 0.25 eV (at 0.25 ~ 0.45 eV). Above 0.45 eV, the Pauli blocking is no longer significant for the two nodes, and substantial compensation of their CPGE contributions occurs.

The above scenario agrees with the DFT-derived $\eta(\omega)$ spectrum for undoped CoSi [16] and applies to both our and earlier experimental observations in the limited frequency ranges [16]. Particularly, the measured spectral hump (Fig. 5) is explained as the high-frequency part of the bipolar CPGE lineshape, which originates from a relevant R−node contribution compensated by an increasing Γ−node effect for $\omega > 0.45$ eV. The expected sign-reversed low-frequency part of the bipolar lineshape is, however, out of our spectral range, despite the measured $\eta(\omega)$ already exhibits a decreasing tendency below 0.38 eV. Note that the explanation relies on $E_F$ presumed near the charge neutral point. We confirm this presumption by dc Hall and magnetoresistance measurements (see SM section S3 for details). The fitting on these field-dependent transport data at 2 K with a two-band model yields an electron (hole) concentration of $1.65\times10^{20}$ cm$^{-3}$ ($1.21\times10^{20}$ cm$^{-3}$), which falls close to previously published data [31,32]. This reveals an electron pocket at the R point and a minor hole pocket at the Γ point, as predicted by the DFT-derived band structure for a stoichiometric crystal. A tiny difference of $E_F$ between our and previously studied samples may explain the different linewidths of their $\eta(\omega)$ spectra [16].

Finally, we comment on three issues. First, the measured lower $\eta(\omega)$ compared to Ref. [16] may arise from different sample conditions, e.g., $E_F$ and/or the scattering rate, but we cannot exclude possible influences of different theoretical modeling for the THz emission. Secondly, the spectral hump at 0.26 ~ 0.43 eV (Fig. 5) is unlikely a signature of the quantized CPGE because of the incomplete optical transitions near the Γ point and the high-order band dispersion of the bottom cone of the R-point node [16]. Thirdly, we have also determined the linear shift conductivity of CoSi through Eq. (7), yielding $\rho \approx 17 \pm 3$ μA/V$^2$ below 0.7 eV. While the value is lower than $\eta(\omega)$, it is readily comparable to benchmark materials in the visible range, such as ferroelectric oxides (~5 μA/V$^2$ for BaTiO$_3$ and PbTiO$_3$ [1,33]) and ferroelectric semiconductor SbSI (~8 μA/V$^2$ [17]), and can thus greatly expands the photovoltaic applications for utility in thermal energy conversion and sensing in the infrared range.

## V. CONCLUSIONS

We characterize the nonlinear photocurrents in chiral multifold CoSi with the improved TES analysis, through which the photon drag effect and the bulk transverse injection current are identified. In the mid-infrared range, we find a large linear shift conductivity, and confirm a giant injection conductivity as a consequence of strongly interfered non-quantized contributions from the vicinity of multifold nodes with opposite chiralities. This study supports the chiral transition metal monosilicides for mid-infrared photovoltaics via various nonlinear optical channels, but also consolidates the THz spectroscopy for quantitative topological nonlinear optics research

**Acknowledgments:** The authors thank Sin-Yi Wei, Tien-Dat Tran, Trieu Huu Tho, and Si-Tong Liu for experimental supports and Dr. Naoki Ogawa for inspiring discussions. This work was funded by Academia Sinica (grant number: AS-iMATE-113-12, AS-iMATE-113-14, AS-iMATE-111-12, and AS-GCS-111-M01), and National Science and Technology Council, Taiwan (grant number: NSTC 113-2112-M-001-055-, NSTC 113-2124-M-001-003, and NSTC 113-2112-M-001-045-MY3) The authors acknowledge resources and supports from the X-Ray Diffraction Materials Analysis Laboratory of the Institute of Physics, Academia Sinica.

**Figure Captions:**

FIG. 1. (a) Schematics of the TES setup and definitions of geometric parameters. Optical pumps excite the sample set at the azimuthal angle $\phi$. The radiated THz field is analyzed with a wire gird polarizer (WGP) through the electro-optic sampling. (b) Illustration of an incident wave in air (medium I) and a transmitted wave in a lossy medium II. The transmitted field has a complex wavevector, whose real and imaginary parts refer to the normals of the planes with constant phase and constant amplitude, respectively.

FIG. 2. Measured THz transients emitted from CoSi in *SS*, *PP*, *PS*, *RP*, and *LP* polarization combinations at $\phi = 0°$ and $\hbar\omega = 0.61$ eV. For each, the optical and THz polarizations are labelled in order as *S*, *P*, *R*, or *L* for *s*-, *p*-, right-hand circular, or left-hand circular polarization, respectively. Data are vertically shifted for clarity, and the horizontal thin lines depict the corresponding zeros.

FIG. 3. Measured amplitude of the THz field emitted from CoSi at $\phi = 0°$ and $\hbar\omega = 0.61$ eV as a function of the rotation angle of a QWP in the beam path of the optical pump (dots). Upper and bottom panels show the results of *p*- and *s*-polarized THz components, respectively. Solid lines are the fitting curves based on Eq. (5), which can be decomposed into the contributions from the PGE (dashed lines) and the PDE plus PDmE (dotted lines). Horizontal thin lines depict zero.

FIG. 4. Measured amplitude of the THz field emitted from CoSi as a function of the azimuthal angle $\phi$ in (a) *SS* and (c) *PS* polarization combinations at $\hbar\omega = 0.61$ eV (dots). Black lines in

(a) and (c) are LPGE-derived fitting curves that depict the $\phi$-dependent experimental features. Red line in (c) is a fit with both LPGE and CPGE taken into account. All the curves in (a) and (c) are calculated with a constant value of $\rho$. Horizontal thin lines depict zeros. Polar plots (b) and (d) represent the same data and fits in the absolute strength.

FIG. 5. Measured circular injection conductivity of CoSi single crystal versus the pump photon energy. Inset shows a schematic of the multifold band crossings at the $\Gamma$ and R points (not on the realistic scales). The bands constituting the degenerate spin-1 fermion at $\Gamma$ are labelled as band 1 to 3 from bottom to top.

**Figures:**

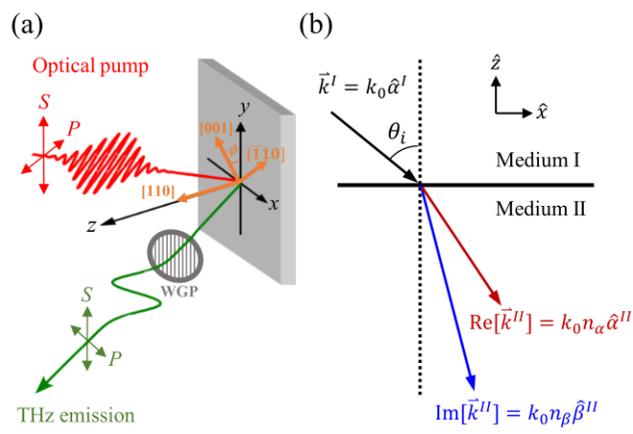

# Figure 1

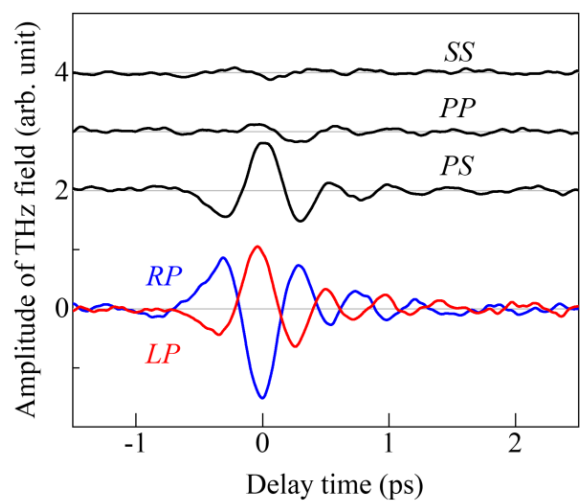

**Figure 2**

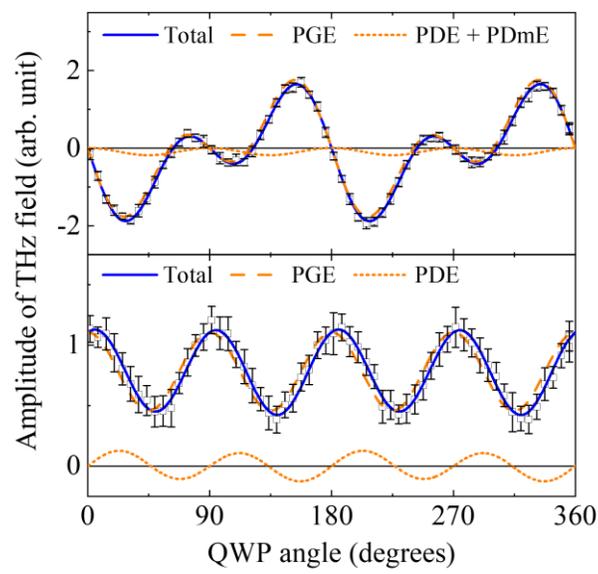

**Figure 3**

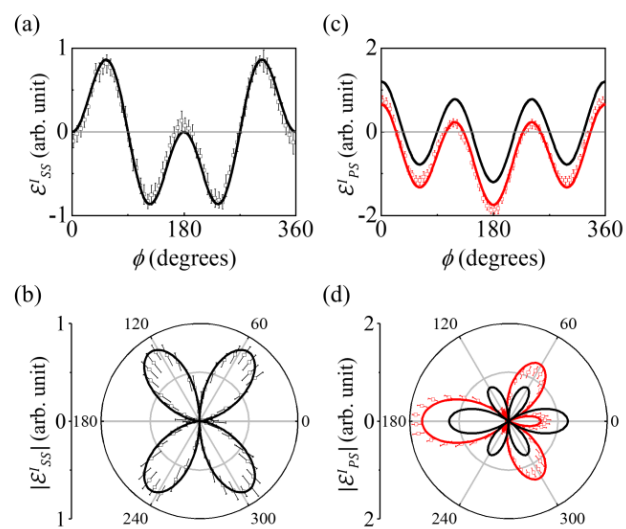

**Figure 4**

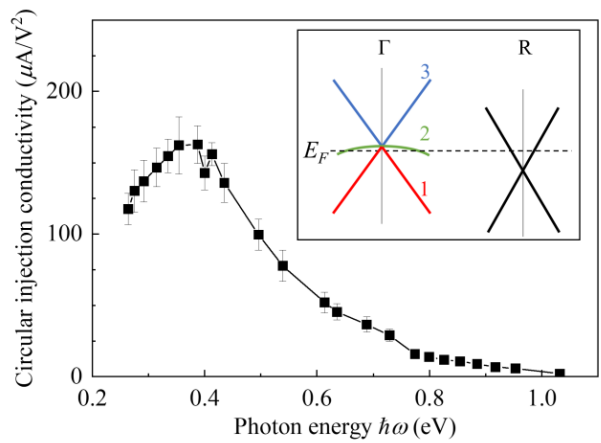

**Figure 5**

Supplementary Material for

# Origin of nonlinear photocurrents in chiral multifold semimetal CoSi unveiled by terahertz emission spectroscopy

**S1. Growth of single crystal and structural characterization**

**S2. Phase mismatch and Fresnel coefficients for the air/CoSi interface**

**S3. Transport measurements and two-band fitting**

**S4. Supplementary equations**

**S1. Growth of single crystal and structural characterization**

High-quality single crystal of CoSi was grown using the modified optical floating zone method. At first, prior to initiating the crystal growth process, the 5N purity of Cobalt (pellet) and Silicon (Si) pieces were taken in the molar ratio of (1:1.1) and directly melted in the arc melting furnace under the Ar atmosphere. To ensure the melt uniformity, the prepared alloy samples were flipped over and re-melted at least three times for each sample. The crystal phase of the as-prepared CoSi alloys was identified as a single phase (without observable impurity phase) using powder X-ray diffraction. Its space group was determined as $P2_13$, with lattice constants a = c = 4.45 Å. To prepare CoSi single crystals, prepared polycrystalline CoSi samples were kept in the bottom conical-shaped Alumina tube and hung to the upper shaft of the optical floating zone furnace, and given the constant heating and rotation for efficient homogeneity melting. We adopted a growth rate of 0.1 mm/hr with a constant Ar flow rate of 0.2 L/min. The grown crystal was cut and polished.

Then, the technique of backscattering Laue diffraction was employed to characterize the crystallographic quality of the fabricated crystal and determine lattice orientation. As shown in Fig. S1, the measured Laue diffraction pattern reveals clear and sharp spots without twinning features, confirming its single-domain crystallinity with (110) surface orientation. This conclusion was also supported by a rotational-anisotropy optical sum-frequency generation (SFG) analysis, as shown and discussed in Fig. S2.

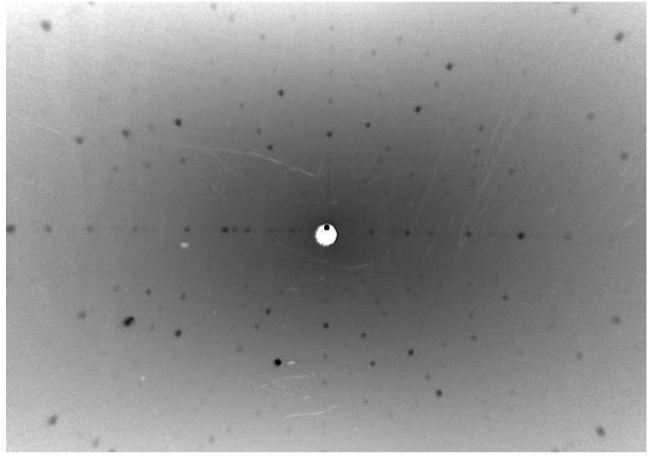

FIG. S1. Backscattering Laue diffraction pattern of (110) CoSi single crystal

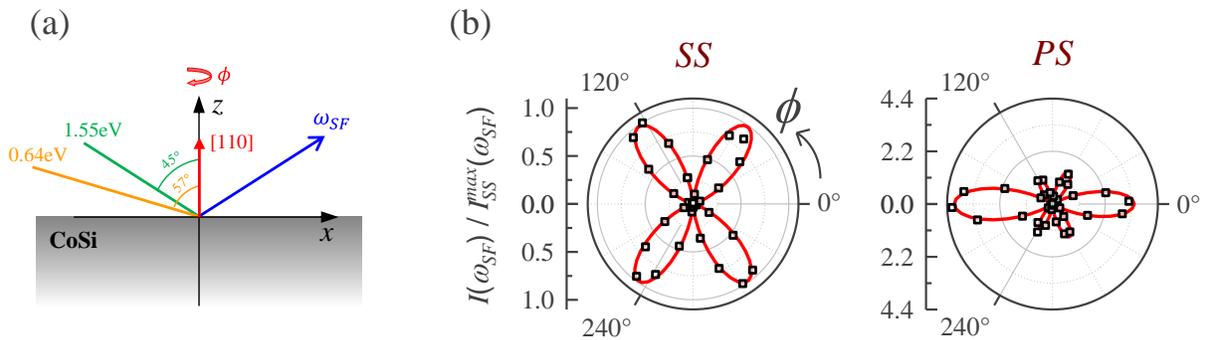

FIG. S2. (a) Schematic of the SFG experiment with azimuthal rotation. Two input beams at 0.64 eV and 1.55 eV have the same polarization. (b) Normalized SFG intensity, $I(\omega_{SF})$, from CoSi as a function of $\phi$ for *SS* and *PS* polarization configurations (dots). Polarization of the input and output fields are labelled in order. Lines are theoretical fits based on $P2_13$ space group of the crystal, as expressed by $I_{SS}(\omega_{SF}) \propto |\cos\phi \sin^2\phi|^2$ and $I_{PS}(\omega_{SF}) \propto |1 + X_1 \cos\phi + X_2(2\cos\phi - 3\cos^3\phi)|^2$ with $X_{1,2}$ the fitting parameters.

## S2. Phase mismatch and Fresnel coefficients for the air/CoSi interface

We follow Refs. [1-3] to derive the transmission Fresnel coefficients at an interface between air and a non-magnetic, isotropic, absorbing medium, and the calculate the phase mismatch for the reflected intra-pulse DFG.

As sketched in Fig. 1(b) in the main text, we consider that a harmonic homogeneous plane wave (HHPW) of light is incident upon the interface from medium I (air, $z > 0$), and the transmitted field appears as a harmonic inhomogeneous plane wave (HIPW) in the absorbing medium II ($z < 0$) due to the boundary conditions. Particularly, the light field at frequency $\omega$ is expressed as $\vec{E}^m(\vec{r},t) = \vec{E}^m(\vec{k}^m, \omega)e^{i(\vec{k}^m \cdot \vec{r} - \omega t)}$ with $m = I, II$ for denoting the two media. In medium II, the complex wavevector of HIPW is described by $\vec{k}^{II} = \vec{\alpha}^{II} + i\vec{\beta}^{II} = k_0(n_\alpha \hat{\alpha}^{II} + in_\beta \hat{\beta}^{II})$. The unit vectors $\hat{\alpha}^{II}$ and $\hat{\beta}^{II}$ depict the normals of the planes with constant phase and constant amplitude, respectively. They do not coincide in lossy media in general. $n_\alpha$ and $n_\beta$ can be interpreted as *apparent* refractive indices, which relate to $\tilde{\epsilon}^{II}$ through $\vec{k}^{II} \cdot \vec{k}^{II} = k_0^2 \tilde{\epsilon}^{II}$ [1-3]. For the HHPW in air, we simply have $\vec{k}^I = k_0 \hat{\alpha}^I$. Note that $\vec{k}^I$ and $\vec{k}^{II}$ are correlated through continuity of the tangential electric fields in the two media at the interface. Together with the relation $\vec{k}^{II} \cdot \vec{k}^{II} = k_0^2 \tilde{\epsilon}^{II}$, one can solve the complex $\vec{k}^{II}$ for a given $\vec{k}^I$, as expressed by [3]

$$\hat{x} \cdot \vec{k}^{II} = k_0 \sin\theta_i, \tag{S1a}$$

$$\hat{y} \cdot \vec{k}^{II} = 0, \tag{S1b}$$

$$\hat{z} \cdot \vec{k}^{II} = -k_0 \left[\frac{\left(s^2+\epsilon_2^{II^2}\right)^{1/2}+s}{2}\right]^{1/2} - ik_0 \left[\frac{\left(s^2+\epsilon_2^{II^2}\right)^{1/2}-s}{2}\right]^{1/2}, \text{ with } s \equiv \epsilon_1^{II} - \sin^2\theta_i, \tag{S1c}$$

where $\tilde{\epsilon}^{II} = \epsilon_1^{II} + i\epsilon_2^{II}$. As a check, one can set $\epsilon_2^{II} = 0$ for non-absorbing materials, and then find the above relations reduce to the standard Snell's law, as expected.

In formulating the Fresnel factors, we are aware that HIPWs cannot be decomposed into the usual *s* and *p* polarizations in general. Here we follow Dupertuis *et al.* [1] to define a set of complex orthonormal polarization unit vectors:

$$\hat{e}_{PE} = \frac{\hat{z} \times \vec{k}}{|\hat{z} \times \vec{k}|} \text{ and } \hat{e}_{PM} = \frac{\vec{k} \times \hat{e}_{PE}}{k_0 \tilde{n}} \tag{S2}$$

with $\tilde{n}$ the complex refractive index, and $\hat{e}_{PE} \cdot \hat{e}_{PM} = 0$. The projection of $\vec{E}$ along $\hat{e}_{PE}$ is termed as parallel electric (PE) mode because $\hat{e}_{PE}$ is always parallel to the plane interface. Similar considerations hold of course for the parallel magnetic (PM) modes. The transmission coefficients for the PE and PM modes across an interface are derived as [1-3]

$$t_{PE} = \frac{2k_z^I}{k_z^I + k_z^{II}} \text{ and } t_{PM} = \frac{2\tilde{n}^I \tilde{n}^{II} k_z^I}{\tilde{n}^{II^2} k_z^I + \tilde{n}^{I^2} k_z^{II}}. \tag{S3}$$

One can relate the components of the incident and transmitted fields by the transmission Fresnel coefficients defined as a diagonal matrix $\overleftrightarrow{L}$ with the relation $\vec{E}^{II}(\vec{k}^{II}, \omega) = \overleftrightarrow{L} \vec{E}^I(\vec{k}^I, \omega)$. Its elements are expressed as

$$L_{xx} \equiv \frac{\hat{x}\cdot\vec{E}^{II}(\vec{k}^{II},\omega)}{\hat{x}\cdot\vec{E}^{I}(\vec{k}^{I},\omega)} = \frac{2k_z^{II}}{\tilde{n}^{II^2}k_z^I+k_z^{II}}, \qquad (S4a)$$

$$L_{yy} \equiv \frac{\hat{y}\cdot\vec{E}^{II}(\vec{k}^{II},\omega)}{\hat{y}\cdot\vec{E}^{I}(\vec{k}^{I},\omega)} = \frac{2k_z^I}{k_z^I+k_z^{II}}, \qquad (S4b)$$

$$L_{zz} \equiv \frac{\hat{z}\cdot\vec{E}^{II}(\vec{k}^{II},\omega)}{\hat{z}\cdot\vec{E}^{I}(\vec{k}^{I},\omega)} = \frac{2k_z^I}{\tilde{n}^{II^2}k_z^I+k_z^{II}}. \qquad (S4c)$$

Note that Eq. (S2) and (S3) also apply to the HHPWs, for which the PE and PM modes reduce to be the usual *s*- and *p*-polarized waves, respectively. Eq. (S4) is identical to the expressions of $\overleftrightarrow{L}$ for HHPW [4], but we now consider $\vec{k}^{II} = \vec{\alpha}^{II} + i\vec{\beta}^{II}$ for HIPWs. Besides the optical field, Eq. (S4) is utilized to relate the THz fields in the two media with its emission angle $\theta_e$ in medium I.

The phase mismatch between the optical and THz fields dictates their interaction length. It is defined as $\Delta k_z^{II} = k_z^{II}(\omega + \omega_{THz}) - k_z^{II}(\omega)^* - k_z^{II}(\omega_{THz})$ for reflected intra-pulse DFG process. Symbol * denotes the complex conjugate. With $\omega \gg \omega_{THz}$, this relation can be simplified to be

$$\Delta k_z^{II}(\omega,\omega_{THz}) = 2i\cdot\text{Im}[k_z^{II}(\omega)] - \frac{\omega_{THz}}{\omega}\cdot\lambda_{op}\cdot\frac{\partial k_z^{II}(\omega)}{\partial\lambda}\bigg|_{\lambda=\lambda_{op}} - k_z^{II}(\omega_{THz}), \qquad (S5)$$

where $\lambda_{op}$ is the optical wavelength.

In calculating $\overleftrightarrow{L}$ and $\Delta k_z^{II}$ for CoSi, we use Eq. (S1) to calculate $\vec{k}^{II}$ for the optical and THz fields with the known $\tilde{\epsilon}^{II}$ [5] and $\theta_i = 45°$, and then use the deduced $\vec{k}^{II}$ to calculate $\overleftrightarrow{L}$ at both optical and THz frequencies and $\Delta k_z^{II}(\omega,\omega_{THz})$ via Eq. (S4) and (S5), respectively. Similar calculations are also done for the ZnTe emitter with its $\tilde{\epsilon}^{II}$ [6,7]. The results are summarized in Fig. S3 and S4.

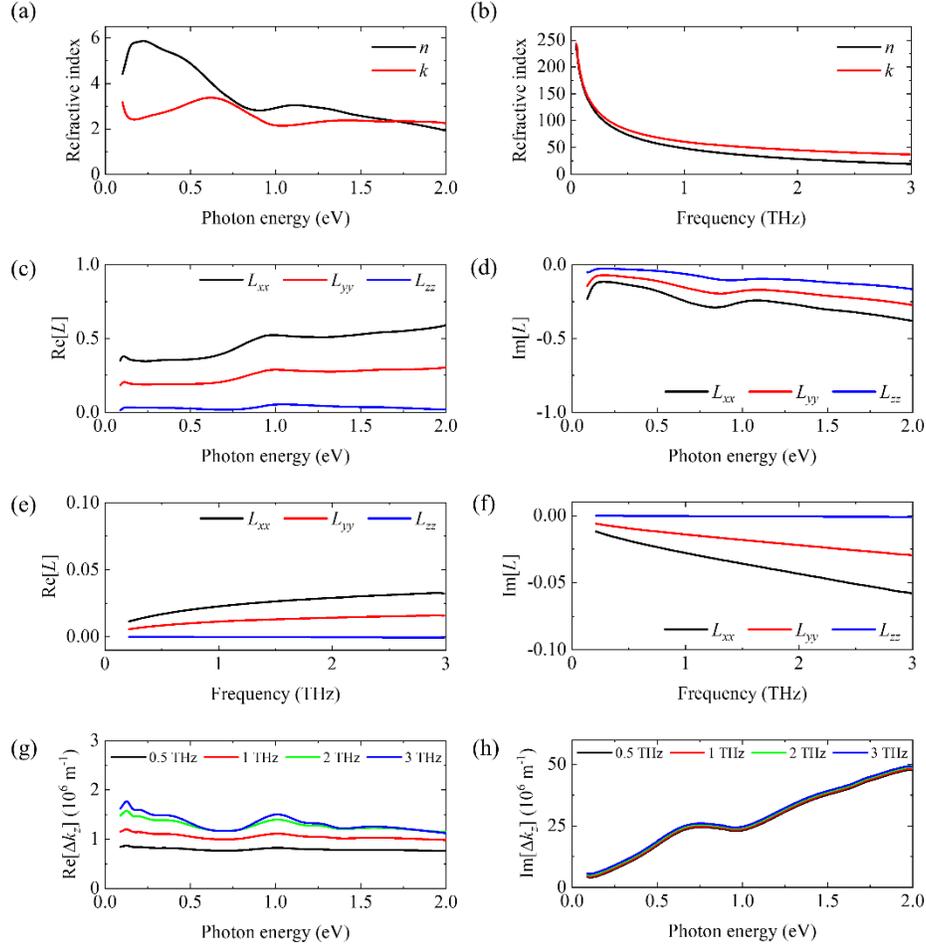

FIG. S3. Calculated $\overleftrightarrow{L}$ and $\Delta k_z^{II}$ for the air/CoSi interface. (a) and (b) complex refractive index $\tilde{n} = n + ik$ of CoSi in the optical and THz regions, respectively [5]. (c) and (d) [(e) and (f)] are the real and imaginary parts of $\overleftrightarrow{L}$ in the optical [THz] region, respectively. (g) and (h) are the real and imaginary parts of $\Delta k_z^{II}(\omega, \omega_{THz})$. Note that $\text{Im}[\Delta k_z^{II}] \gg \text{Re}[\Delta k_z^{II}]$ reflects the fact that the interaction length of the waves in CoSi is essentially limited by their absorption depths.

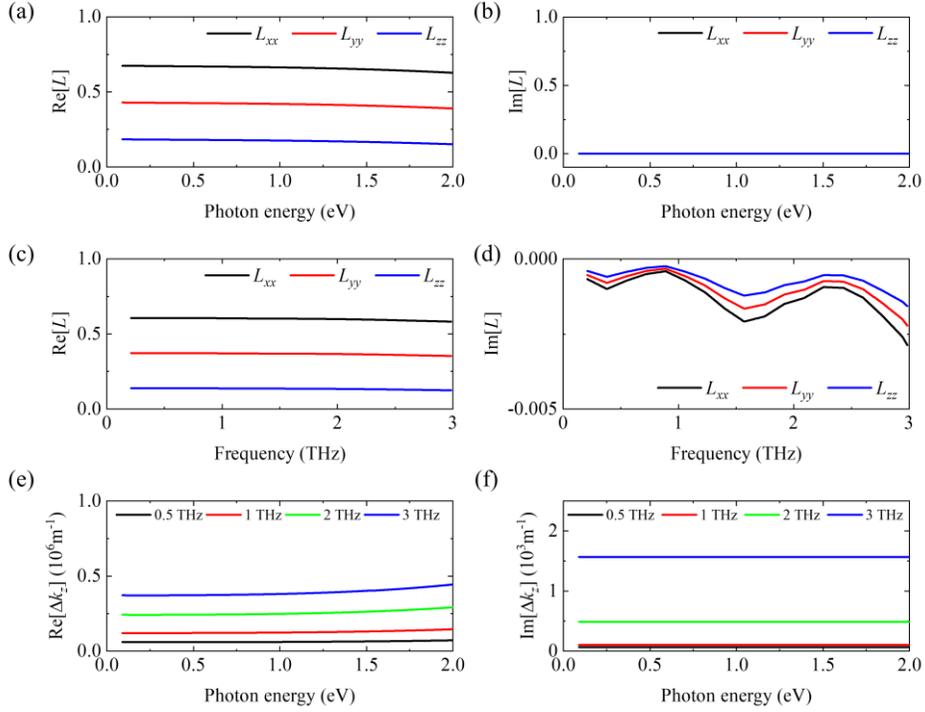

FIG. S4. Calculated $\overleftrightarrow{L}$ and $\Delta k_z^{II}$ for the air/ZnTe interface. (a) and (b) [(c) and (d)] are the real and imaginary parts of $\overleftrightarrow{L}$ in the optical [THz] region, respectively. (e) and (f) are the real and imaginary parts of $\Delta k_z^{II}(\omega, \omega_{THz})$. In (b), $\mathrm{Im}\overleftrightarrow{L}$ is essentially zero in the optical region below the band gap.

## S3. Transport measurements and two-band fitting

To determine the carrier density of the CoSi single crystal, we conducted the measurements of Hall conductance ($\sigma_{xy}$) and magnetoconductance ($\sigma$) with the magnetic field along [110] at $T = 2$ K, as shown in Fig. S5 (a) and (b), respectively. We extracted the carrier's type and their densities using the standard two-band model fitting to the experimental $\sigma_{xy}$ data. In the two-band model [8-10], the $\sigma_{xy}$ and $\sigma$ can be expressed as

$$\sigma_{xy}(B) = eB \left[ \frac{n_1 \mu_1^2 s_1}{1+(\mu_1 B)^2} + \frac{n_2 \mu_2^2 s_2}{1+(\mu_2 B)^2} \right] \text{ and} \tag{S6}$$

$$\sigma(B) = e \left[ \frac{n_1 \mu_1}{1+(\mu_1 B)^2} + \frac{n_2 \mu_2}{1+(\mu_2 B)^2} \right], \tag{S7}$$

where $e$ is the electronic charge, and $B$ is the magnetic field. The density, associated mobility, and sign of carrier type for the band 1 (band 2) are denoted by $n_1$ ($n_2$), $\mu_1$ ($\mu_2$), and $s_1$ ($s_2$), respectively, where $s = +1$ and $-1$ are assigned to be hole-type and electron-type, respectively. An additional constraint from the measured zero-field conductivity of $\sigma_0 = en_1\mu_1 + en_2\mu_2$ was introduced to the fitting of $\sigma_{xy}(B)$ data using Eq. (S6). The resulting fitting curve [red-dashed line in in Fig. S5 (a)] with two opposite signs of carrier type turns out to give excellent fitting to the experimental $\sigma_{xy}(B)$ data. The extracted two band parameters are $(n_1, \mu_1, s_1) = (1.21 \times 10^{20} \text{ cm}^{-3}, 127 \text{ cm}^2\text{V}^{-1}\text{s}^{-1}, +1)$ and $(n_2, \mu_2, s_2) = (1.65 \times 10^{20} \text{ cm}^{-3}, 402 \text{ cm}^2\text{V}^{-1}\text{s}^{-1}, -1)$. For consistent check, the calculated $\sigma(B)$ [red-dashed line in Fig. S5(b)] using the same extracted two band parameters and Eq. (S7) shows relatively good agreement with the experimental data, which thus supports for the validity of our two-band analysis. We note that our extracted two band parameters are in accord with the previously published data and also DFT band structure calculations for a stoichiometric crystal [10,11], revealing an electron pocket at the R point and a hole pocket at the $\Gamma$ point.

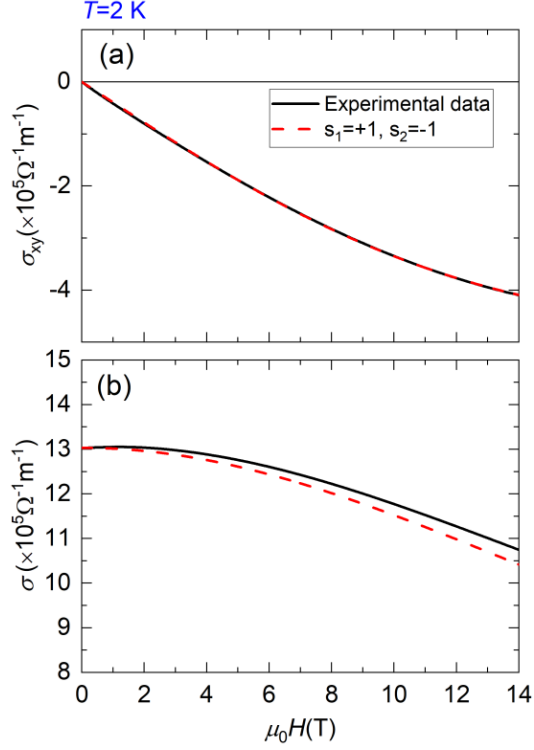

FIG. S5. Two-band model fitting of the magnetotransport data of CoSi. (a) The black-solid and red-dashed line represent experimental data and the two-band model fitting curve of $\sigma_{xy}(B)$ at 2 K, respectively, by considering opposite signs of carrier type ($s_1 = +1$ and $s_2 = -1$) for the two band. (b) The extracted two band parameters from the $\sigma_{xy}(B)$ fitting were used to calculate the corresponding $\sigma(B)$ curve at 2 K. The calculated $\sigma(B)$, represented by the red-dashed line, shows reasonable agreement with the experimental data.

**S4. Supplementary equations**

For CoSi of space group 198, there are five independent non-zero PDE tensor elements: $v^{(2)}_{aaaa} = v^{(2)}_{bbbb} = v^{(2)}_{cccc}$, $v^{(2)}_{bccb} = v^{(2)}_{caac} = v^{(2)}_{abba}$, $v^{(2)}_{cbbc} = v^{(2)}_{acca} = v^{(2)}_{baab}$, $v^{(2)}_{bbcc} = v^{(2)}_{ccaa} = v^{(2)}_{aabb} = v^{(2)*}_{bcbc} = v^{(2)*}_{caca} = v^{(2)*}_{abab}$, and $v^{(2)}_{ccbb} = v^{(2)}_{aacc} = v^{(2)}_{bbaa} = v^{(2)*}_{cbcb} = v^{(2)*}_{acac} = v^{(2)*}_{baba}$. Together with the PGE, the constituted photoconductivity tensor $\overleftrightarrow{\sigma}^{(2)}$ in the $(a,b,c)$ crystal coordinates is expressed by Eq. (4) in the main text and represented here in the (*x*,*y*,*z*) laboratory coordinates by a rotation transformation via $\sigma_{ijk}^{(2)} = \sum_{lmn} R_{il} R_{jm} R_{kn} \sigma_{lmn}^{(2)}$ with $i, j, k \in (x,y,z)$ and $l, m, n \in (a,b,c)$. For $\hat{z} \parallel [110]$ and $\hat{y} \parallel [001]$, we have the rotation transformation tensor $\overset{\leftrightarrow}{R} = \frac{1}{\sqrt{2}}\begin{pmatrix} 1 & -1 & 0 \\ 0 & 0 & -\sqrt{2} \\ 1 & 1 & 0 \end{pmatrix}$ and deduce $\sigma_{ijk}^{(2)}$ as

$$\vec{\sigma}^{(2)} = 2\sqrt{2} \begin{pmatrix} \begin{pmatrix} P_3 + P_7 - P_2 \\ -2\sqrt{2}\rho \\ P_8 + P_4 - P_1 \end{pmatrix} & \begin{pmatrix} -2\sqrt{2}\rho \\ P_{11} \\ i2\sqrt{2}\eta \end{pmatrix} & \begin{pmatrix} -P_4 + P_8 - P_1 \\ -2\sqrt{2}i\eta \\ -P_3 + P_7 - P_2 \end{pmatrix} \\ \begin{pmatrix} -2\sqrt{2}\rho \\ P_{10}^* \\ -2\sqrt{2}i\eta \end{pmatrix} & \begin{pmatrix} P_{10} \\ 0 \\ P_9 \end{pmatrix} & \begin{pmatrix} i2\sqrt{2}\eta \\ P_9^* \\ 2\sqrt{2}\rho \end{pmatrix} \\ \begin{pmatrix} P_8 - P_6 + P_1 \\ i2\sqrt{2}\eta \\ P_5 + P_7 + P_2 \end{pmatrix} & \begin{pmatrix} -i2\sqrt{2}\eta \\ P_{12} \\ 2\sqrt{2}\rho \end{pmatrix} & \begin{pmatrix} -P_5 + P_7 + P_2 \\ 2\sqrt{2}\rho \\ P_8 + P_6 + P_1 \end{pmatrix} \end{pmatrix}, \text{with} \quad (S8)$$

$P_1 = \alpha_a v_{bccb}^{(2)} + \alpha_b v_{cbbc}^{(2)}$,

$P_2 = \alpha_a v_{bccb}^{(2)} - \alpha_b v_{cbbc}^{(2)}$,

$P_3 = 2\alpha_b \text{Re} v_{bbcc}^{(2)} - 2\alpha_a \text{Re} v_{ccbb}^{(2)}$,

$P_4 = 2i\alpha_a \text{Im} v_{ccbb}^{(2)} + 2i\alpha_b \text{Im} v_{bbcc}^{(2)}$,

$P_5 = 2i\alpha_a \text{Im} v_{ccbb}^{(2)} - 2i\alpha_b \text{Im} v_{bbcc}^{(2)}$,

$P_6 = 2\alpha_a \text{Re} v_{ccbb}^{(2)} + 2\alpha_b \text{Re} v_{bbcc}^{(2)}$,

$P_7 = (\alpha_b - \alpha_a) v_{aaaa}^{(2)}$,

$P_8 = (\alpha_a + \alpha_b) v_{aaaa}^{(2)}$,

$P_9 = 2\left(\alpha_b v_{ccbb}^{(2)} + \alpha_a v_{bbcc}^{(2)}\right)$,

$P_{10} = 2\left(\alpha_b v_{ccbb}^{(2)} - \alpha_a v_{bbcc}^{(2)}\right)$,

$$P_{11} = 2\left(\alpha_b v^{(2)}_{bccb} - \alpha_a v^{(2)}_{cbbc}\right),$$

$$P_{12} = 2\left(\alpha_a v^{(2)}_{cbbc} + \alpha_b v^{(2)}_{bccb}\right),$$

where the element $\sigma^{(2)}_{ijk}$ is the *k*-th element of the column vector in the *i*-th row and *j*-th column of the matrix.

In describing the TES experiment with a rotating QWP (Fig. 3 in the main text), we consider that an optical pulse linearly polarized on the scattering plane travels through a QWP with its fast axis at angle $\psi$ away from the scattering plane before pumping the sample. The polarization unit vector of the pump field is thus expressed by

$$\hat{e}^I(\omega) = e^{-\frac{i\pi}{4}} \begin{pmatrix} (\cos^2\psi + i\sin^2\psi)\cos\theta_i \\ (1-i)\sin\psi\cos\psi \\ (\cos^2\psi + i\sin^2\psi)\sin\theta_i \end{pmatrix}. \tag{S9}$$

Here $\psi = 0°, 45°$, and $135°$ correspond to *p*-polarized, left-hand circularly polarized, and right-hand circularly polarized pumps, respectively. We express the amplitude of the emitted THz field from CoSi through Eq. (2) in the main text and Eq. (S8) with the $\psi$-dependent $\hat{e}^I(\omega)$. For the *s*-polarized THz emission, it is expressed as $\mathcal{E}^I_S(\omega_{THz}) \propto A_1 \cdot \cos 4\psi + A_2 \cdot \sin 4\psi + A_3 \cdot \sin 2\psi + A_4$, with

$$A_1 = \rho[|L_{zz}(\omega)|^2 - |L_{xx}(\omega)|^2] - 2\eta\,\text{Im}[L_{zz}(\omega)L^*_{xx}(\omega)], \tag{S10a}$$

$$A_2 = \tfrac{1}{4}\alpha_a \left\{ \text{Re} v_{bbcc} \left[ \text{Re}\left(L_{zz}(\omega)L^*_{yy}(\omega)\right) - \text{Re}\left(L_{xx}(\omega)L^*_{yy}(\omega)\right) \right] + \text{Im} v_{bbcc} \left[ \text{Im}\left(L_{xx}(\omega)L^*_{yy}(\omega)\right) - \text{Im}\left(L_{zz}(\omega)L^*_{yy}(\omega)\right) \right] \right\} + \tfrac{1}{4}\alpha_b \left\{ \text{Re} v_{ccbb} \left[ \text{Re}\left(L_{xx}(\omega)L^*_{yy}(\omega)\right) + \text{Re}\left(L_{zz}(\omega)L^*_{yy}(\omega)\right) \right] - \text{Im} v_{ccbb} \left[ \text{Im}\left(L_{xx}(\omega)L^*_{yy}(\omega)\right) + \text{Im}\left(L_{zz}(\omega)L^*_{yy}(\omega)\right) \right] \right\},$$
(S10b)

$$A_3 = \tfrac{1}{2}\alpha_a \left\{ \text{Re} v_{bbcc} \left[ \text{Im}\left(L_{xx}(\omega)L^*_{yy}(\omega)\right) - \text{Im}\left(L_{zz}(\omega)L^*_{yy}(\omega)\right) \right] + \text{Im} v_{bbcc} \left[ \text{Re}\left(L_{xx}(\omega)L^*_{yy}(\omega)\right) - \text{Re}\left(L_{zz}(\omega)L^*_{yy}(\omega)\right) \right] \right\} - \tfrac{1}{2}\alpha_b \left\{ \text{Re} v_{ccbb} \left[ \text{Im}\left(L_{xx}(\omega)L^*_{yy}(\omega)\right) + \text{Im}\left(L_{zz}(\omega)L^*_{yy}(\omega)\right) \right] + \text{Im} v_{ccbb} \left[ \text{Re}\left(L_{xx}(\omega)L^*_{yy}(\omega)\right) + \text{Re}\left(L_{zz}(\omega)L^*_{yy}(\omega)\right) \right] \right\},$$
(S10c)

$$A_4 = 3A_1.$$
(S10d)

Clearly, $A_2$ and $A_3$ ($A_1$ and $A_4$) originate from the PDE (PGE). Similarly, we derive $\mathcal{E}^I_P(\omega_{THz}) \propto A_1 \cdot \cos 4\psi + A_2 \cdot \sin 4\psi + A_3 \cdot \sin 2\psi + A_4$ for $p$-polarized THz radiations, with

$$A_1 = \left(L_{zz}(\omega_{THz})\sigma^{(2)}_{zxx} - L_{xx}(\omega_{THz})\sigma^{(2)}_{xxx}\right)|L_{xx}(\omega)|^2 + \left(L_{zz}(\omega_{THz})\sigma^{(2)}_{zxz} - L_{xx}(\omega_{THz})\sigma^{(2)}_{xxz}\right)L_{xx}(\omega)L^*_{zz}(\omega) + \left(L_{zz}(\omega_{THz})\sigma^{(2)}_{zzx} - L_{xx}(\omega_{THz})\sigma^{(2)}_{xzx}\right)L_{zz}(\omega)L^*_{xx}(\omega) + \left(L_{zz}(\omega_{THz})\sigma^{(2)}_{zzz} - L_{xx}(\omega_{THz})\sigma^{(2)}_{xzz}\right)|L_{zz}(\omega)|^2 + 2\left(L_{xx}(\omega_{THz})\sigma^{(2)}_{xyy} - L_{zz}(\omega_{THz})\sigma^{(2)}_{zyy}\right)|L_{yy}(\omega)|^2 + \text{DP}_1,$$
(S11a)

$$A_2 = 8\left[\rho\left(L_{xx}(\omega_{THz})\text{Re}\left(L_{xx}(\omega)L^*_{yy}(\omega)\right) + L_{zz}(\omega_{THz})\text{Re}\left(L_{yy}(\omega)L^*_{zz}(\omega)\right)\right) + \eta\left(L_{xx}(\omega_{THz})\text{Im}(L_{yy}(\omega)L^*_{zz}) - L_{zz}(\omega_{THz})\text{Im}\left(L_{xx}(\omega)L^*_{yy}(\omega)\right)\right)\right],$$
(S11b)

$$A_3 = -16\left[\rho\left(L_{xx}(\omega_{THz})\text{Im}\left(L_{xx}(\omega)L^*_{yy}(\omega)\right) - L_{zz}(\omega_{THz})\text{Im}(L_{yy}(\omega)L^*_{zz})\right) + \eta\left(L_{xx}(\omega_{THz})\text{Re}\left(L_{yy}(\omega)L^*_{zz}(\omega)\right) + L_{zz}(\omega_{THz})\text{Re}\left(L_{xx}(\omega)L^*_{yy}(\omega)\right)\right)\right],$$
(S11c)

$$A_4 = 3\left(L_{zz}(\omega_{THz})\sigma^{(2)}_{zxx} - L_{xx}(\omega_{THz})\sigma^{(2)}_{xxx}\right)|L_{xx}(\omega)|^2 + 3\left(L_{zz}(\omega_{THz})\sigma^{(2)}_{zxz} - L_{xx}(\omega_{THz})\sigma^{(2)}_{xxz}\right)L_{xx}(\omega)L^*_{zz}(\omega) +$$

$$3\left(L_{zz}(\omega_{THz})\sigma^{(2)}_{zzx} - L_{xx}(\omega_{THz})\sigma^{(2)}_{xzx}\right)L_{zz}(\omega)L^*_{xx}(\omega) + 3\left(L_{zz}(\omega_{THz})\sigma^{(2)}_{zzz} - L_{xx}(\omega_{THz})\sigma^{(2)}_{xzz}\right)|L_{zz}(\omega)|^2 +$$

$$2\left(L_{zz}(\omega_{THz})\sigma^{(2)}_{zyy} - L_{xx}(\omega_{THz})\sigma^{(2)}_{xyy}\right)|L_{yy}(\omega)|^2 + DP_4. \tag{S11d}$$

It is found that the PGE and PDE contribute to $A_2 \cdot \sin 4\psi + A_3 \cdot \sin 2\psi$ and $A_1 \cdot \cos 4\psi + A_4$ separately. (Here we avoid lengthy expressions of $A_1$ and $A_4$ by keeping the PDE-derived $\sigma^{(2)}_{lmn}$ terms.) In addition, we consider that the PDmE follows a $\psi$-dependent optical reflectance of the sample to contribute to $A_1$ and $A_4$, as denoted by extra $DP_1$ and $DP_4$ terms in Eq. (S11a) and (S11d).

Finally, in measuring the PGE coefficients of CoSi, we measure the THz radiations with *PS*, *LP*, and *RP* polarization combinations at $\phi = 0°$, and normalize these spectra against the ZnTe reference emitter. To analyze the THz spectra, we express the corresponding THz field amplitudes from Eq. (2) in the main text and Eq. (S8), as given by

$$\mathcal{E}^I_{PS}(\omega_{THz}) = (D_1 \cdot \rho + D_2 \cdot \eta)|\mathcal{E}^I(\omega)|^2, \tag{S12a}$$

$$\Delta\mathcal{E}^I(\omega_{THz}) \equiv (\mathcal{E}^I_{LP} - \mathcal{E}^I_{RP})/2 = (D_3 \cdot \rho + D_4 \cdot \eta)|\mathcal{E}^I(\omega)|^2, \tag{S12b}$$

$$\mathcal{E}^{I,ZT}(\omega_{THz}) = \chi^{ZT} D_0 |\mathcal{E}^I(\omega)|^2, \tag{S12c}$$

with

$$D_1 \equiv \frac{f L_{yy}(\omega_{THz})}{2\omega_{THz}\Delta k_z^{II}}[|L_{xx}(\omega)|^2 - |L_{zz}(\omega)|^2], \tag{S12d}$$

$$D_2 \equiv \frac{f L_{yy}(\omega_{THz})}{\omega_{THz}\Delta k_z^{II}}\operatorname{Im}[L_{zz}(\omega)L^*_{xx}(\omega)], \tag{S12e}$$

$$D_3 \equiv \frac{f}{2\omega_{THz}\Delta k_z^{II}}\left[L_{xx}(\omega_{THz})\operatorname{Im}[L_{xx}(\omega)L^*_{yy}(\omega)] + L_{zz}(\omega_{THz})\operatorname{Im}[L_{zz}(\omega)L^*_{yy}(\omega)]\right], \tag{S12f}$$

$$D_4 \equiv \frac{f}{2\omega_{THz}\Delta k_z^{II}}\left[L_{zz}(\omega_{THz})\text{Re}[L_{xx}(\omega)L_{yy}^*(\omega)] + L_{xx}(\omega_{THz})\text{Re}[L_{yy}(\omega)L_{zz}^*(\omega)]\right], \quad \text{(S12g)}$$

$$D_0 \equiv f\frac{i\epsilon_0 \sin\phi}{2\sqrt{2}\Delta k_z^{II}} \cdot \{L_{xx}(\omega_{THz})[3\cos^2\phi|L_{xx}(\omega)|^2 - |L_{zz}(\omega)|^2] + 2L_{zz}(\omega_{THz})L_{xx}(\omega)L_{zz}(\omega)\}.$$

(S12h)

For ZnTe, we have considered $\overleftrightarrow{\sigma}^{(2)} = -i\omega_{THz}\epsilon_0 \cdot \overleftrightarrow{\chi}^{(2)} = -i\omega_{THz}\epsilon_0 \cdot \chi^{ZT}|\epsilon_{lmn}|$. Its azimuthal angle $\phi$ is defined as that for CoSi and set to yield the optimized optical rectification (~36°). We calculate $D_i$ ($i = 0\sim4$) from Eq. (S12) using $\overleftrightarrow{L}$ and $\Delta k_z^{II}$ shown in Fig. S3 and S4.